\documentclass[conference, final, 10pt]{IEEEtran}

\usepackage{cite}
\usepackage{graphicx}
\usepackage{array}
\usepackage{amsmath, amssymb}
\usepackage{bm}
\usepackage{textpos}
\usepackage{subfigure}
\usepackage{enumerate}
\usepackage{lscape}
\usepackage{float}
\usepackage{pdfsync}
\usepackage{url}
\usepackage{fancybox}
\usepackage{ragged2e}
\usepackage{color}
\usepackage{mathdots}

\usepackage[acronym,shortcuts]{glossaries}
\newacronym{AI}{AI}{artificial intelligence}
\newacronym{AR}{AR}{augmented reality}
\newacronym{AWGN}{AWGN}{additive white Gaussian noise}
\newacronym{BS}{BS}{base station}
\newacronym{BER}{BER}{bit-error rate}
\newacronym{CCMC}{CCMC}{continuous-input-continuous-output memoryless channel}
\newacronym{CSI}{CSI}{channel state information}
\newacronym{CSIT}{CSIT}{CSI at the transmitter}
\newacronym{CV}{CV}{connected vehicles}
\newacronym{DCMC}{DCMC}{discrete-input-continuous-output memoryless channel}
\newacronym{DFT}{DFT}{discrete Fourier transform}
\newacronym{GSD}{GSD}{generalized sphere decoder}
\newacronym{IOT}{IoT}{Internet of Things}
\newacronym{IIOT}{IIoT}{industrial IoT}
\newacronym{mmWave}{mmWave}{millimeter wave}
\newacronym{MIMO}{MIMO}{multiple-input multiple-output}
\newacronym{MUI}{MUI}{mutual interference}
\newacronym{OFDM}{OFDM}{orthogonal frequency division multiplexing}
\newacronym{QAM}{QAM}{quadrature amplitude modulation}
\newacronym{PHY}{PHY}{physical layer}
\newacronym{RE}{RE}{resource element}
\newacronym{RF}{RF}{radio frequency}
\newacronym{SISO}{SISO}{single-input single-output}
\newacronym{SDR}{SDR}{software-defined radio}
\newacronym{UE}{UE}{user equipment}
\newacronym{V2X}{V2X}{vehicle-to-everything}
\newacronym{VR}{VR}{virtual reality}
\newacronym{LTE}{LTE}{long-term evolution}
\newacronym{FD}{FD}{full duplex}

\newacronym{CDMA}{CDMA}{code division multiple access}
\newacronym{CDNOMA}{CD-NOMA}{code-domain NOMA}
\newacronym{IDMA}{IDMA}{interleave division multiple access}
\newacronym{IGMA}{IGMA}{interleave-grid multiple access}
\newacronym{LDS}{LDS}{low-density spreading}
\newacronym{LPMA}{LPMA}{lattice partition multiple access}
\newacronym{MCNOMA}{MC-NOMA}{massively concurrent NOMA}
\newacronym{MUSA}{MUSA}{multi-user shared access}
\newacronym{NOMA}{NOMA}{non-orthogonal multiple access}
\newacronym{NCMA}{NCMA}{non-orthogonal coded multiple access}
\newacronym{NOCA}{NOCA}{non-orthogonal coded access}
\newacronym{OMA}{OMA}{orthogonal multiple access}
\newacronym{PDMA}{PDMA}{pattern division multiple access}
\newacronym{PDNOMA}{PD-NOMA}{power-domain NOMA}
\newacronym{RDMA}{RDMA}{repetition division multiple access}
\newacronym{RSMA}{RSMA}{resource spread multiple access}
\newacronym{SCMA}{SCMA}{sparse-coded multiple access}
\newacronym{WSMA}{WSMA}{Welch-bound spreading multiple access}

\newacronym{CSIDCO}{CSIDCO}{complex successive iterative decorrelation by convex optimization}
\newacronym{ETF}{ETF}{equiangular tight frames}
\newacronym{FP}{FP}{frame potential}
\newacronym{UNTF}{UNTF}{unit-norm tight frame}
\newacronym{SSC}{SSC}{sum of squared correlations}
\newacronym{WB}{WB}{Welch Bound}

\newacronym{BP}{BP}{belief propagation}
\newacronym{ESE}{ESE}{elementary signal estimator}
\newacronym{GaBP}{GaBP}{Gaussian belief propagation}
\newacronym{ML}{ML}{maximum likelihood}
\newacronym{MMSE}{MMSE}{minimum mean squared error}
\newacronym{MMSESIC}{MMSE-SIC}{MMSE-SIC}
\newacronym{MMSEPIC}{MMSE-PIC}{MMSE-PIC}
\newacronym{MPA}{MPA}{message passing algorithm}
\newacronym{MUD}{MUD}{multi-user detection}
\newacronym{PIC}{PIC}{parallel interference cancellation}
\newacronym{SIC}{SIC}{successive interference cancellation}

\newacronym{eMBB}{eMBB}{enhanced mobile broadband}
\newacronym{URLLC}{URLLC}{ultra reliable low-latency communications}
\newacronym{mMTC}{mMTC}{massive machine-type communications}
\newacronym{2G}{2G}{second generation}
\newacronym{3G}{3G}{third generation}
\newacronym{4G}{4G}{fourth generation}
\newacronym{5G}{5G}{fifth generation}
\newacronym{6G}{6G}{sixth generation}
\newacronym{NR}{NR}{new radio}

\newcommand{\bv}[1]{\mathbf #1}

\title{6G: the Wireless Communications Network for Collaborative and AI Applications\\[-1ex]}
\author{
\IEEEauthorblockN{Razvan-Andrei Stoica, and Giuseppe Thadeu Freitas de Abreu}
\IEEEauthorblockA{Focus Area Mobility, Jacobs University Bremen, Campus Ring 1, 28759, Bremen, Germany \\
Emails: {\tt [r.stoica,g.abreu]@jacobs-university.de}}\\[-5ex]
}

\begin{document}

\maketitle

\begin{abstract}
At the dawn of \acs{5G}, we take a leap forward and present an original vision of wireless communication beyond its horizon towards \acs{6G}:
a paradigm-shifting perspective of wireless networks on the cusp of an \acs{AI} revolution.
\end{abstract}

\glsresetall

\vspace{-1ex}
\section{Introduction}
\label{sect:intro}
\vspace{-1ex}

In recent years, Academia and Industry alike converged to a \ac{NR} specification referred to as \ac{5G}, whose standard was just released in June 2018 \cite{etsi2018rel15_5g}.
A freeze of the aforementioned standard will only happen, however, around the second quarter of 2019, with earliest deployments of \ac{5G} networks expected for late 2019.
It goes without saying that such deployments will not fully implement all features envisioned by the \ac{5G} standard, but a mere subset of these.
Subsequent releases and updates are expected to follow under what is being referred to (somewhat unimaginatively) as \ac{5G} \ac{LTE}.

The hype about \ac{5G}, not only in Academia and Industry, but also in the Media is well-justified by the promised gains in terms of rate, accessibility and reliability of wireless services, and respectively, the shift in the architecture model of \ac{5G} as opposed to its \ac{4G} predecessor and legacy systems.
Concretely, the improvements of \ac{5G} over its \ac{4G} predecessor are mainly due to a paradigm shift in the design architecture of wireless systems, which under \ac{5G} is ground-up, aimed at solving real business requirements \cite{etsi20185garticle}.

Amongst such design patterns of \ac{5G} is the replacement of the hierarchical layered network architecture of the past into a more flexible format, capable of functional virtualization and network aggregation.
Another major deviation from old established ways was the resolution to move beyond the canonical sub-$6$ Ghz bands of preceding systems towards \acp{mmWave}, leading to a tenfold increase of available bandwidth.
From a connectivity density perspective, \ac{5G} (in its later versions) will break beyond carrier aggregation, moving decisively towards non-orthogonal concurrent access in both uplink and downlink \cite{wu2018comprehensivestudynoma5g}.
These advances are finally coupled with great progress in \ac{RF} hardware, going beyond \ac{MIMO} to add, for the first time in history, wireless full-duplex capabilities.

In the light of all the above, one could ask: ``\emph{What else could we want? Why even bother with thinking of \ac{6G}?}''.
The answer to these rhetorical questions is that, as usual, revolution never comes from within, but is rather imposed by radical changes in exterior conditions.
And that radical change, which is now beaming straight towards the wireless communication world ready to cause major disruption, is the raise of \ac{AI}.

%
%

\section{\hspace{-0.3ex}\acs{AI}\hspace{-0.1ex}: Paradigm\hspace{-0.2ex} shift\hspace{-0.2ex} for\hspace{-0.2ex} Wireless\hspace{-0.2ex} Communications}
\label{sect:ai}

One of the most important advances in the history of the man-kind is currently blossoming in the form of \ac{AI} \cite{russell2016artificial}.
Computational intelligence bears the prospects of a trendsetting technology able to unlock solutions to previously difficult and large-scale problems outside of the current cloud-centric paradigm.
Intelligent agents trained in the cloud using machine learning algorithms on Big Data will be deployed in the real world in the next decades.
Such entities will be tasked to solve multiple optimization problems across a vast set of business verticals, empowering new business models and industries alike and leading to a technological revolution.

But in order to harness the true power of such agents, \textit{collaborative \ac{AI}} is the key.
And by nature of the mobile society of the 21st century, it is clear that this collaboration can only be achieved via wireless communications.

The proliferation of sensors in modern day appliances, coupled with the aforementioned advances will lead to \textit{advanced context-awareness} which can be collaboratively leveraged towards common goals.
Consider for instance a fleet of autonomous vehicles (the moving \aclp{BS} of the future) driving collaboratively through a crowded urban canopy, freed of traffic lights / signs.
The collaboration of intelligent agents (in this case the autonomous vehicles) will be mostly locally-oriented (connectivity amongst vehicles nearing a given crossing takes precedence), but augmented with outer layers (the planned routes of each vehicle, the prospective proximity of emergency vehicles \textit{etc.}) that need to be taken into account.

Under the requirements of dynamic applications, designed on the fly by \ac{AI} nodes sharing a common goal -- such as the one illustrated above -- \textit{collaborative quilt networks} will be generated bottom-up.
Interactions will therefore be necessary in vast amounts, to solve large distributed problems where massive connectivity, large data volumes and ultra low-latency beyond those to be offered by \ac{5G} networks will be essential.

But, the intelligent agents can be deployed to \acp{BS} as well, where \ac{AI} technology can be leveraged not only for network optimization tasks, a direction which \ac{5G} network slicing and virtualization is aiming to solve, but also to provide business and application intelligence for users.
%
%

%
In light of the above arguments some prospective qualitative requirements of a future \ac{6G} standard are:
\begin{itemize}
 \item functional/situational/positional network \textit{self-aggregation}
 \item pervasive \textit{enhanced context-awareness}
 \item network / nodes \textit{contextual self-reconfiguration} 
 \item \textit{opportunistic} latency, rate and access setup
\end{itemize}
%

%
In summary, we envision that the emergence of \ac{AI} will be the driver of \ac{6G}, which will enable the proliferation of \textit{distributed independent autonomous systems} and \textit{associated common goal-driven massive fog-computing clusters}.

\vspace{-1ex}
\section{The Role of \acs{PHY} Layer in 6G}
\label{sect:phy}

It is clear from the above list of \ac{6G} requirements that the \textit{policy-driven} network slicing and virtualization at the heart of \ac{5G} will not be sufficiently adaptive to address the needs of the future.
The \ac{AI}-central \textit{context-aware adaptive device self-reconfiguration} and \textit{network self-aggregation} envisioned at the core of \ac{6G} will require thus vast research for its modular development and seamless integration into the canonically rigid communications ecosystem.

%

But one would be mistaken to think that \ac{6G} will focus on network aspects, and that the PHY layer in \ac{6G} will have a secondary role.
In fact, the research community was at a similar cross-roads at the dawn of \ac{4G}, when it asked itself if ``\emph{the PHY layer was dead}'' \cite{dohler2011isthephydead}.
The concern then was that \ac{PHY} research would slowly fade and limit to satisfying short-term and derivative needs of the Industry.
Those concerns were short-lived, as proven by the powerful revival propelled by ground-breaking innovations such as \ac{NOMA}, \ac{FD} radio and hybrid \ac{mmWave} \ac{RF} technologies.

Learning from this recent history, we believe that \ac{PHY} research will not become extinct but will transform and solve future challenges along the road, as discussed in the prequel.
Concretely, we believe that \ac{PHY} will be separated in two components: the \textit{Low-\ac{PHY}} and the \textit{High-\ac{PHY}}.

The first will focus on solving signal processing problems close to hardware, including the handling of waveforms, beamforming \& interference management, all taking into account the hardware impairments imposed by the compacting of massive MIMO antennas, full-duplex operation and \ac{mmWave} \ac{RF}, as well as the imperfections of \ac{CSI}, caused by all the dynamics of \ac{6G} architectures.
This ``\emph{Low-PHY}'' will also develop strongly towards enabling \acp{SDR}, a dream many times attempted before, but which never really blossomed.

In turn, the ``\emph{High-PHY}'' will move towards becoming the \textit{software oriented driver} of the Low-\ac{PHY}.
Research on High-PHY will therefore focus on the interface with the \ac{AI} core, developing the code-domain technologies required to control and interact with the Low-\ac{PHY}.

To cite an example, take the last requirement (opportunism) cited at the end of Section \ref{sect:ai}.
Although \ac{5G} aims to deliver low-latency and high connectivity density, it will still employ the traditional model of grant-based access \cite{etsi2018rel15_5g}.
In contrast, a truly opportunistic system, requires the implementation of a grant-free access scheme, such as the one proposed in \ac{MCNOMA} \cite{stoica2018mcnomaasilomar}, in which \textit{non-orthogonal massively concurrent} access is enabled both at downlink and uplink thanks to the exploitation of properties of \textit{Frame Theory} \cite{casazza2012finite}.
\ac{MCNOMA} can seamlessly multiplex $K$ active users onto $M < K$ available orthogonal resources, \textit{e.g.} time slots, frequency tones, spatial streams, spreading the symbols of active users onto \textit{all} available resources.

As an example, the \ac{MCNOMA} scheme transmission over a group of \acs{OFDM} sub-carrriers leades to the following received signal model corresponding to a $k$-th user
\vspace{-0.5ex}
\begin{equation}
\label{eq:system_downlink}
\bm{y}_k = \bv{H}_{k,\rm DL}\bv{F}\cdot\bm{s} + \bm{n}_k,
\end{equation}
where the \textit{frame} $\bv{F} \in \mathbb{C}^{M\times K}$ forms an \textit{overcomplete virtual signature waveform dictionary} that carefully spreads \textit{all} the active users' symbols $\bm{s} \in \mathbb{C}^K$ upon the available pool of $M$ subtones transmitted over the downlink fading channel modeled by the diagonal matrix $\bv{H}_{k, \rm DL}$.

From the above, it is easy to foresee how \ac{MCNOMA} can be expanded also to include duty-cycles (periods of activity and inactivity), such that the method can be a basis to completely eliminate the need for network discovery procedures.

\begin{figure}[H] 
 \centering
 \includegraphics[width=0.46\textwidth]{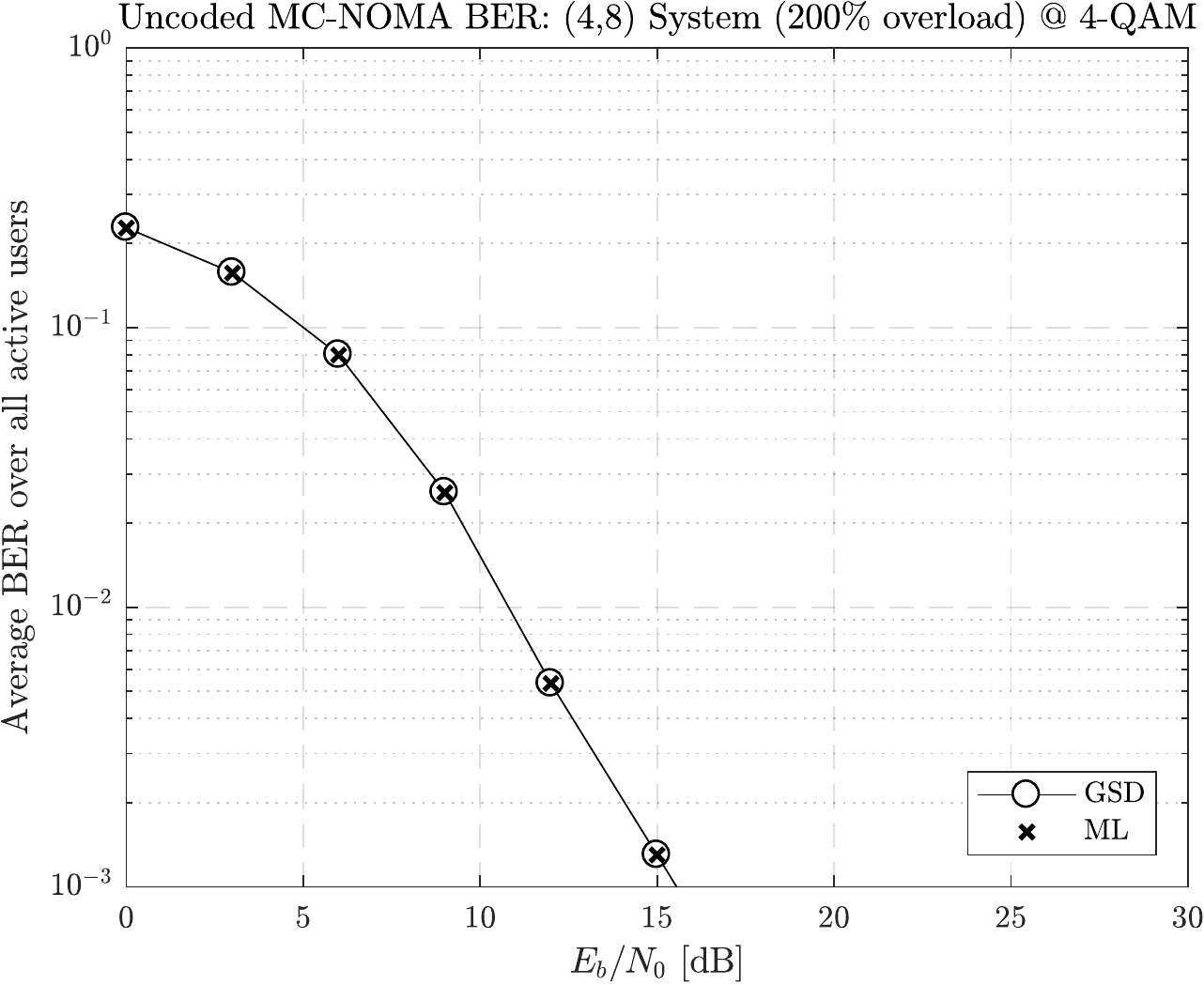}
 \vspace{-2ex}
 \caption{BER performance over Rayleigh fading at 200\% overloading.}\label{fig:4x8ber}
 \vspace{-1ex}
\end{figure}

To be complete, an \ac{MCNOMA} scheme must include not only the optimum design of $\bv{F}$, but also a practical \ac{MUD} at the receiver.
These problems are solved using \textit{randomized incoherent tight frames}, coupled with efficient low-complexity stochastic \acp{GSD} at the receiver, leading to optimum \ac{ML} detection \cite{MyListOfPapers:AndreiMCNOMA_Journal2019}, as illustrated in Fig \ref{fig:4x8ber}.
\hspace{-2ex}


\end{document}